\documentclass{iaus}
\usepackage{graphicx}
\usepackage[dvips]{epsfig}
\title[HE~1327$-$2326, a dwarf or subgiant with $\mbox{[Fe/H]}=-5.4$] {The new
record holder for the most iron-poor star: HE~1327$-$2326, a dwarf or subgiant
with $\mbox{[Fe/H]}=-5.4$}

\author[Frebel et al.]  
{A. Frebel$^1$
\break W. Aoki$^2$, N. Christlieb$^3$, H. Ando$^2$, M. Asplund$^1$,
P. S. Barklem$^4$, T. C. Beers$^5$, K. Eriksson$^4$, C. Fechner$^3$,
M. Y. Fujimoto$^6$, S. Honda$^2$, T. Kajino$^2$, T. Minezaki$^7$,
K. Nomoto$^8$, J. E. Norris$^1$, S. G. Ryan$^9$, M. Takada-Hidai$^{10}$,
S. Tsangarides$^9$ \and Y. Yoshii$^7$}

\affiliation{$^1$Research School of Astronomy \& Astrophysics, Australian
   National University, Australia \break email:
   anna@mso.anu.edu.au\\
[\affilskip]$^2$ National Astronomical Observatory of Japan, Japan\\
[\affilskip]$^3$ Hamburger Sternwarte, Germany\\
[\affilskip]$^4$ Department of Physics \& Space Sciences, Uppsala
  Astronomical Observatory, Sweden\\
[\affilskip]$^5$ Department of Physics \& Astronomy, and JINA: Joint
  Institute for Nuclear Astrophysics, Michigan State University, USA\\
[\affilskip]$^6$ Department of Physics, Hokkaido University, Japan\\
[\affilskip]$^7$ Institute of Astronomy, School of Science, University of
  Tokyo, Japan\\
[\affilskip]$^8$ Department of Astronomy, School of Science, University of
  Tokyo, Japan\\
[\affilskip]$^9$ Department of Physics and Astronomy, Open University, UK\\
[\affilskip]$^{10}$ Liberal Arts Education Center, Tokai University, Japan}

\pubyear{2005}
\volume{228}  
\pagerange{1--7}
\date{?? and in revised form ??}
\jname{From Lithium to Uranium: Elemental Tracers of Early Cosmic Evolution}
\editors{V. Hill, P. Fran\c{c}ois \& F. Primas, eds.}
\begin{document}

\maketitle

\begin{abstract}
We describe the discovery of HE~1327$-$2326, a dwarf or subgiant with
$\mbox{[Fe/H]}=-5.4$. The star was found in a sample of bright metal-poor
stars selected from the Hamburg/ESO survey. Its abundance pattern is
characterized by very high C and N abundances. The detection of Sr which is
overabundant by a factor of 10 as compared to iron and the Sun, suggests that
neutron-capture elements had already been produced in the very early Galaxy. A
puzzling Li depletion is observed in this unevolved star which contradicts the
value of the primordial Li derived from WMAP and other Li studies.  Possible
scenarios for the origin of the abundance pattern (Pop. II or Pop. III) are
presented as well as an outlook on future observations.

\end{abstract}

\firstsection 
\section{Introduction}
Observations of metal-poor stars with $\mbox{[Fe/H]}<-3.0$ provide crucial
clues to the formation of the first objects in the Universe and their
contribution to its chemical enrichment. The detailed abundance analyses of
such stars reveal observational details which are compared to theoretical
models for nucleosynthesis yields originating from e.g. first generation
supernovae, in order to learn about the chemical evolution of the Galaxy.

Until recently only one star was known to have an iron abundance with
$\mbox{[Fe/H]}<-4.0$ (HE~0107$-$5240 with $\mbox{[Fe/H]}=-5.2$; Christlieb et
al. 2002). Now, a second star, HE~1327$-$2326, has been found with
$\mbox{[Fe/H]}=-5.4$ (Frebel et al. 2005, Aoki et al. in preparation). These
objects provide insight into the early Universe because the existence of the
two low-mass stars challenges the current scenarios for the formation of the
very first stars after the Big Bang. The abundance patterns of both stars pose
many questions to observers and theoreticians and much progress has been made
in order to explain how these very particular objects could form. This paper
describes the very recent discovery of the new record holder for the most
iron-poor object, HE~1327$-$2326.

\section{Bright metal-poor stars from the Hamburg/ESO Survey}
Thus far, the Hamburg/ESO (HES) survey was only investigated for its fainter
metal-poor stars (Christlieb et al. 2003, Beers et al., this volume). However,
it was recently extended to the brighter end ($10<B<14$; Frebel et al. in
preparation). Despite partial saturation effects, it was possible to select a
sample of $1777$ bright metal-poor candidate stars.

Three observational steps are necessary to identify the most metal-poor
stars. Figure 1 illustrates these steps.
\begin{figure}
 \begin{center}
  \epsfig{file=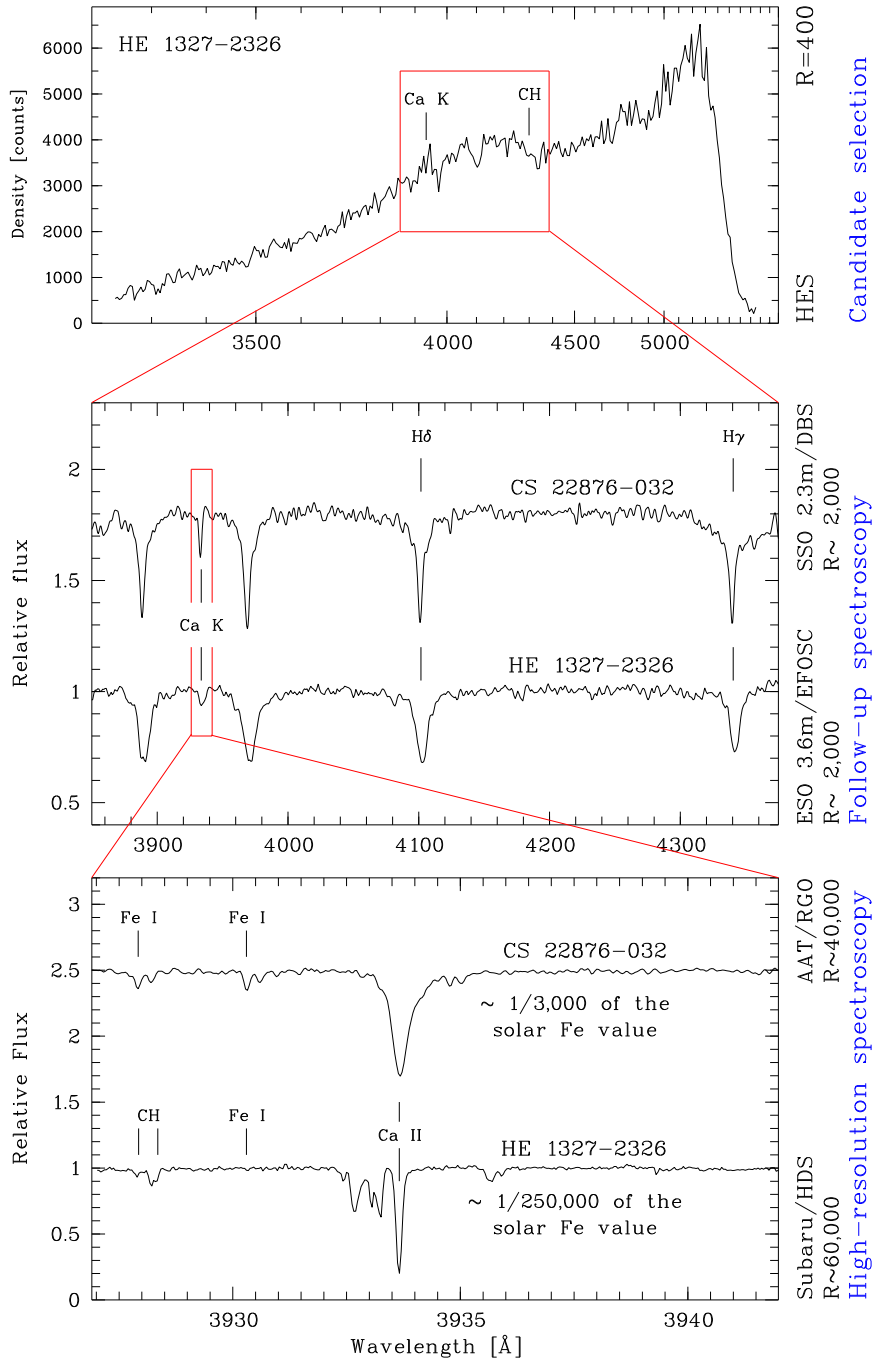, clip=, width=11.5cm, bbllx=83,
  bblly=39, bburx=547, bbury=769}
  \caption{The three observational steps to find metal-poor stars illustrated
  by means of HE~1327$-$2326. Top panel: HES objective-prism spectrum. Middle
  panel: Medium-resolution spectrum of HE~1327$-$2326 in comparison with
  CS~22876$-$032 ($\mbox{[Fe/H]}=-3.7$; Norris et al. 2000 and references
  therein). From this data we measured $\mbox{[Fe/H]}=-4.3$ for HE~1327$-$2326
  because interstellar Ca blended with the Ca II K line. Bottom panel:
  High-resolution spectra of both objects. Only with the high-resolution data
  was it possible to determine the true iron abundance, $\mbox{[Fe/H]}=-5.4$,
  for HE~1327$-$2326.}
 \end{center}
\end{figure}
Medium-resolution follow-up spectroscopy ($\sim2$\,{\AA}) of the entire sample
of bright stars was recently completed. In April 2003, a medium-resolution
spectrum of HE~1327$-$2326 was taken with the ESO $3.6$\,m telescope. Using
the Ca II K line at $3933$\,{\AA} and the Beers et al. (1999) calibration we
derived an iron abundance of $\mbox{[Fe/H]}=-4.3$. High-resolution spectra of
HE~1327$-$2326 with the Japanese Subaru telescope and its High Dispersion
Spectrograph were taken in May and June 2004 (see Aoki et al., this
volume). It was immediately revealed that the metallicity estimate obtained
from the medium-resolution data was greatly overestimated due to the presence
of interstellar Ca II. The interstellar feature was not resolved at lower
resolution, thus blending with the stellar Ca II K line. Reddening of
$E(B-V)=0.08$ (Schlegel et al. 1998) in the line-of-sight is consistent with
the presence of the interstellar Ca II (see bottom panel of Figure 1).

Further high-resolution spectroscopy of the most metal-poor stars from the
bright sample is currently underway.

\section{Stellar parameters of HE~1327$-$2326}
From $BVRIK$ photometry obtained with the 2\,m Magnum telescope (University of
Tokyo) we derive an effective temperature of $T_{\mbox{\scriptsize
eff}}=6180\pm80$\,K, based on the Alonso et al. (1996) scale. This temperature
is consistent with the values derived from a Balmer line profile analysis.

Due to the absence of Fe II lines in the spectrum of HE~1327$-$2326, the
gravity could not be determined in a spectroscopic fashion. Both neutral and
ionized species are needed to derive the gravity from the ionization
equilibrium. One Ca I and two Ca II lines are detected in
HE~1327$-$2326. Hence, Ca could in principle be used instead of iron. Using 1D
MARCS and Kurucz model atmospheres, the large difference of 0.8\,dex between
the Ca LTE abundances derived from Ca I and Ca II lines suggests that
significant NLTE effects are present. Thus, we refrained from using Ca for the
gravity determination. Instead, from the known proper motion of HE~1327$-$2326
we inferred an upper limit on its distance and hence its absolute visual
magnitude. This information leads to two different solutions for the gravity
when an isochrone for $\mbox{[Fe/H]}=-3.5$ (Kim et al. 2002) is employed. It
follows that HE~1327$-$2326 is a dwarf ($\log g=4.5$) or a subgiant ($\log g=
3.7$). There is no significant difference in the abundances derived for both
cases. The astrophysical implications are not effected by this uncertainty.

Regarding the iron abundance, we applied a $+0.2$\,dex NLTE correction
(Asplund 2005) resulting in $\mbox{[Fe/H]}=-5.4$ for HE~1327$-$2326.

\section{Summary of the Abundances and Radial Velocity Measurements}

HE~1327$-$2326 has an exceptionally low iron abundance of $\mbox{[Fe/H]}=-5.4$
in combination with extremely high $\mbox{[C/Fe]}$ and $\mbox{[N/Fe]}$ ($\sim
4$\,dex overabundance). See Table 1 for the abundances derived for the dwarf
and the subgiant case. $\mbox{[Mg, Na, Al/Fe]}$ are enhanced by more than one
dex while $\mbox{[Ca, Ti/Fe]}$ are only slightly overabundant. Surprisingly,
the only detected neutron-capture element, Sr, is enhanced by a factor of 10
as compared to iron and the Sun. Li is depleted in this unevolved star and
only an upper limit of $\log \epsilon (Li)=1.6$ could be inferred. This
significantly contradicts the results from other Li studies using unevolved
metal-poor stars (Ryan et al. 1999) as well as the recent WMAP results which
is much higher (by $\sim 1$\,dex; Coc et al. 2004). There is no obvious reason
why the Li is depleted. Possible ideas are that HE~1327$-$2326 might perhaps
be a fast rotator or a member of a binary system. However, no significant line
broadening was found amongst the 17 weak absorption lines used for the
abundance analysis.

\begin{table}
\begin{center}
\caption{\label{abund}Abundance pattern of HE~1327$-$2326 as reported in
Frebel et al. 2005}
 \begin{tabular}{lrr}\hline
Element &\multicolumn{2}{r}\mbox{[Element/Fe]}, A(Li), \mbox{[Fe/H]}\\
        &Subgiant        &Dwarf\\\hline
Li          & $<1.6       $ & $<1.6       $\\
C           & $4.1 \pm 0.2$ & $3.9 \pm 0.2$ \\
N           & $4.5 \pm 0.2$ & $4.2 \pm 0.2$ \\
O           & $<4.0       $ & $<3.7      $  \\
Na (LTE)    & $2.4 \pm 0.2$ & $2.4 \pm 0.2$ \\
Na (non-LTE)& $2.0 \pm 0.2$ & $2.0 \pm 0.2$ \\
Mg (LTE)    & $1.7 \pm 0.2$ & $1.7 \pm 0.2$ \\
Mg (non-LTE)& $1.6 \pm 0.2$ & $1.6 \pm 0.2$ \\
Al (LTE)    & $1.3 \pm 0.2$ & $1.3 \pm 0.2$ \\
Al (non-LTE)& $1.7 \pm 0.2$ & $1.7 \pm 0.2$ \\
Ca I        & $0.1 \pm 0.2$ & $0.1 \pm 0.2$ \\
Ca II       & $0.9 \pm 0.2$ & $0.8 \pm 0.2$ \\
Ti          & $0.6 \pm 0.2$ & $0.8 \pm 0.2$ \\
Fe (LTE)    & $-5.6 \pm 0.2$& $-5.7 \pm0.2$ \\
Fe (non-LTE)& $-5.4 \pm 0.2$& $-5.5 \pm0.2$ \\
Sr (LTE)    & $1.0 \pm 0.2$ & $1.2 \pm 0.2$ \\
Sr (non-LTE)& $1.1 \pm 0.2 $& $1.3 \pm 0.2$ \\
Ba (LTE)    & $<1.4	   $& $<1.7      $  \\	
Ba (non-LTE)& $<1.4       $ & $<1.7$        \\	 \hline
\end{tabular}
\end{center}
\end{table}

\begin{figure}
 \begin{center}
  \epsfig{file=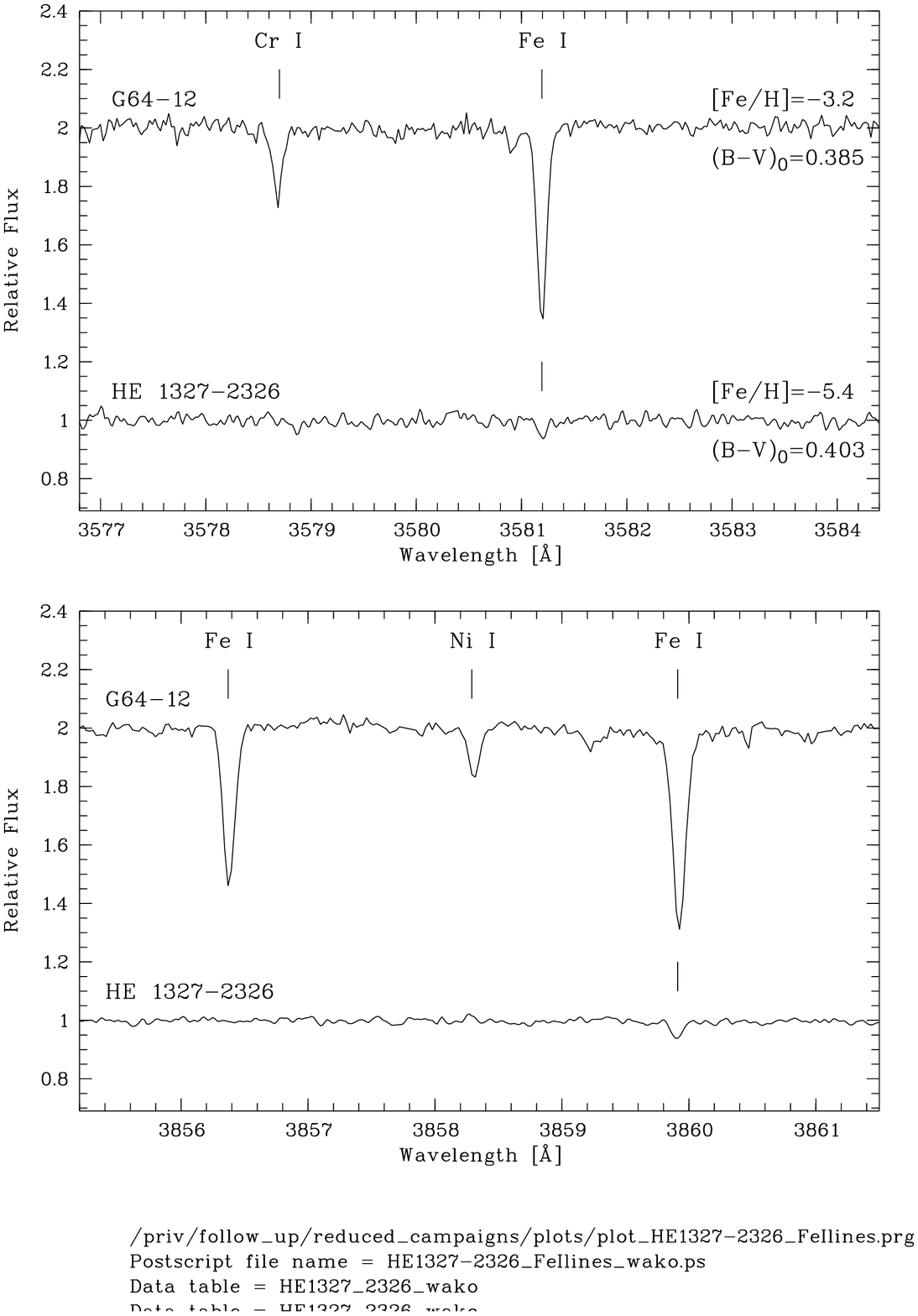, clip=, width=9.6 cm, bbllx=39,
  bblly=109, bburx=541, bbury=769}
  \caption{The two stronger Fe I lines detected in the spectrum of
  HE~1327$-$2326. The lines at 3581\,{\AA} has 5.9\,m{\AA} equivalent width,
  while the line at 3859\,{\AA} has 6.8\,m{\AA}. In total, we detect four Fe I
  lines. The other two have equivalent width of 2.5\,m{\AA} and 1.9\,m{\AA}
  respectively.}  
 \end{center}
\end{figure}

In order to test the binary scenario radial velocities have been determined
from high-resolution observation for four different epochs. All values agree
within their measurement errors which are $\sim1$\,km/s. The results are
summarized in Table 2.
\begin{table}
\begin{center}
\caption{Radial velocity measurements of HE~1327$-$2326 from
  high-resolution data obtained with Subaru/HDS}
 \begin{tabular}{rlll}\hline
& UT date & $v_{rad}$ [km/s] \\\hline 
2004& May 30     & 63.9 & \\
2004& June 02    & 64.0 & \\
2004& June 27    & 63.6 & \\
2005& February 27 & 63.8 & \\ \hline
\end{tabular}
\end{center}
\end{table}
The apparently constant radial velocity has a variety of implications for the
origin of the abundance pattern of HE~1327$-$2326 because it indicates no
membership of a binary system. However, it might still be possible that the
star is a long-period, low-amplitude binary. This has been previously
suggested to be the case for HE~0107$-$5240 (Christlieb et al. 2004b; Suda et
al. 2004). A binary scenario for HE~1327$-$2326 would be able to explain the
Li deficiency. To clarify the situation, radial velocity monitoring of
HE~1327$-$2326 and HE~0107$-$5240 is underway.

The observed Sr/Ba ratio of $\mbox{[Sr/Ba]}>-0.4$ in HE~1327$-$2326 is
inconsistent with that found in s-process-rich stars which have
$\mbox{[Sr/Ba]}<-1.0$. Hence, mass transfer of s-process elements from a
possible former companion seems to be ruled out. The upper limit of the Sr/Ba
ratio is, however, consistent with the ratio found in strongly r-process
enhanced stars (e.g. Christlieb et al. 2004a). This is supportive of an
abundance pattern originating from pre-enrichment by previous generation
supernovae under the assumption that the main site of the r-process is in
supernovae.

\section{Possible Origins of the Abundance Pattern}
There are currently two main different scenarios with which the abundance
pattern of HE~1327$-$2326 might be explained. First, there is a Population II
scenario which invokes pre-enrichment by a previous generation star by means
of a stellar wind or a supernova explosion (Iwamoto et al. 2005; Nomoto et al.,
this volume; Meynet et al. 2005, Meynet et al., this volume). Another
possibility is that HE~1327$-$2326 is a population III star, which accreted
the heavier elements observed on its surface from the interstellar
medium. Lighter elements, such as CNO, would have been donated from a former
AGB companion through mass transfer. The assumption that the star would be
long-period, low amplitude binary is a crucial ingredient.

\section{Conclusions and Outlook}
Further work is required to determine the origin of the abundance patterns of
HE~1327$-$2326 and HE~0107$-$5240. One crucial model-ingredient will be the
oxygen abundance of HE~1327$-$2326. In particular it will challenge the
Population II scenario of Iwamoto et al. and simultaneously provide a test on
the AGB scenario. Hence, 21\,h of VLT/UVES time have recently been granted to
attempt the detection of OH features in the UV.

Clearly, more objects are needed to enlarge the sample of extremely
iron-deficient stars in order to study the phenomena occurring at that low
metallicity range. Using the HES, efforts are underway to observe further
metal-poor candidates as well as to process additional 51 HES plates to
produce even more candidates which might eventually lead to the discovery of
more stars with $\mbox{[Fe/H]}<-5.0$.

\begin{acknowledgments}
Based (in part) on data collected at the Subaru Telescope, which is operated
by the National Astronomical Observatory of Japan.  A.F. thanks the IAU for
financial support to attend this meeting. A.F., M.A. and J.E.N acknowledge
support from the Australian Research Council (grant DP0342613). N.C. is
supported by Deutsche Forschungsgemeinschaft under grants Ch 214/3-1 and Re
353/44-2. P.S.B. and K.E. are supported by the Swedish Research
Council. T.C.B. acknowledges support from grants AST 00-98508, AST 00-98549,
AST 04-06784, and PHY 02-16783, Physics Frontier Centers/JINA: Joint Institute
for Nuclear Astrophysics, awarded by the US National Science Foundation.

\end{acknowledgments}

\end{document}